\documentclass[twocolumn,nofootinbib,aps,superscriptaddress,preprintnumbers,floatfix]{revtex4}

\usepackage{graphicx}
\usepackage{amsmath}
\usepackage[hypertex]{hyperref}

\DeclareMathOperator{\sgn}{sgn}
\DeclareMathOperator{\Li}{Li}

\DeclareMathOperator{\Repart}{Re}

\renewcommand{\Re}{\Repart}

\newcommand{\nn}{\nonumber}
\newcommand{\beq}{\begin{equation}}
\newcommand{\eeq}{\end{equation}}
\newcommand{\beqa}{\begin{eqnarray}}
\newcommand{\eeqa}{\end{eqnarray}}

\renewcommand{\d}{ \mathrm{d} }
\newcommand{\GeV}{ {\,\mathrm{GeV}} }
\newcommand{\ord}{ \mathcal{O} }

\def\OMIT#1{}%

\newcommand{\AFB}{ A_\mathrm{FB} }
\newcommand{\ellpm}{ \ell^+ \ell^- }
\newcommand{\mbbar}{\overline m_b}

\newcommand{\C}{ \mathcal{C} }
\newcommand{\Cs}{ C_7^\mathrm{incl}{} }
\newcommand{\Css}{ \lvert C_7^\mathrm{incl} \rvert^2 }
\newcommand{\Cn}{ C_9^\mathrm{incl} }
\newcommand{\Csn}{ C_{7,9}^\mathrm{incl} }
\newcommand{\Cns}{ \lvert C_9^\mathrm{incl} \rvert^2 }
\newcommand{\ReCsCn}{ \Re(C_7^{\mathrm{incl}*}C_9^\mathrm{incl}) }

\begin{document}

\preprint{
\vbox{
\hbox{hep-ph/0612156}
\hbox{Caltech MAP-330}
\hbox{LBNL-62077}
\hbox{MIT--CTP 3793}
}}

\vspace*{0ex}

\title{\boldmath Extracting short distance information from $b\to s\,\ell^+\ell^-$ effectively}

\author{Keith S.\ M.\ Lee}
\affiliation{California Institute of Technology, Pasadena, CA 91125}

\author{Zoltan Ligeti}
\affiliation{Ernest Orlando Lawrence Berkeley National Laboratory,
University of California, Berkeley, CA 94720}

\author{Iain W.\ Stewart}
\affiliation{Center for Theoretical Physics, Massachusetts Institute of
Technology, Cambridge, MA 02139}

\author{Frank J.\ Tackmann}
\affiliation{Ernest Orlando Lawrence Berkeley National Laboratory,
University of California, Berkeley, CA 94720}


\allowdisplaybreaks[4]

\begin{abstract}

We point out that in inclusive $B\to X_s\ellpm$ decay an angular decomposition
provides a third\linebreak[1] ($q^2$ dependent) observable sensitive to a
different combination of Wilson coefficients than the rate and the
forward-backward asymmetry. Since a precise measurement of $q^2$ dependence
requires large data sets, it is important to consider the data integrated over
regions of $q^2$.  We develop a strategy to extract all measurable Wilson
coefficients in $B\to X_s\ellpm$ from a few simple integrated rates in the low
$q^2$ region. A similar decomposition in $B\to K^*\ellpm$, together with the
$B\to K^*\gamma$ rate, also provides a determination of the Wilson coefficients,
without reliance on form factor models and without having to measure the zero of
the forward-backward asymmetry.

\end{abstract}

\maketitle

\section{Introduction}
\label{sec:intro}

At the scale of $B$ meson decays, flavor changing interactions at the
electroweak scale and above are encoded in Wilson coefficients of operators of
dimension five and higher. The main goal of the $B$ physics program is to make
overconstraining measurements of the magnitudes and phases of these
coefficients~\cite{Hocker:2006xb}, and thereby search for deviations from the
standard model (SM).

The $b\to s\,\ellpm$ process has been observed both in inclusive $B\to X_s
\ellpm$~\cite{Aubert:2004it,Iwasaki:2005sy} and exclusive  $B\to K^{(*)}
\ellpm$~\cite{Abe:2004ir,Aubert:2006vb} decays.  Throughout this paper we assume
the SM except where explicitly stated otherwise.  We also neglect the strange-quark
and lepton masses.  Two observables that have been extensively discussed
for inclusive $B\to X_s\ellpm$ decay are the
$q^2$ spectrum~\cite{Grinstein:1988me} and the forward-backward
asymmetry~\cite{Ali:1991is} (or, equivalently, the energy
asymmetry~\cite{Cho:1996we}).  At lowest order, they are
\begin{align}\label{dGdq2}
\frac{\d\Gamma}{\d q^2}
&= \Gamma_0\,m_b^3\, (1 - s)^2 \Big[
  \big( |C_9|^2 + C_{10}^2 \big) (1 + 2s) \nn\\
& \quad + \frac{4}{s}\, |C_7|^2 (2 + s) + 12\, {\rm Re}(C_7^* C_9) \Big] \,,\\
\frac{\d \AFB}{\d q^2}
&= \int_{-1}^1 \d z\, \frac{\d^2\Gamma}{\d q^2\, \d z}\, \sgn(z) \nn\\
&= -3 \Gamma_0\,m_b^3\, (1 - s)^2\, s\, C_{10}\, {\rm Re}\Big( C_9 + \frac2s C_7\Big)
.\end{align}
Here $q^2 = (p_{\ell^+} + p_{\ell^-})^2$ is the dilepton invariant mass, $s =
q^2/m_b^2$, $z = \cos\theta$, and
\begin{equation}
\Gamma_0 = \frac{G_F^2}{48 \pi^3}\, \frac{\alpha_\mathrm{em}^2}{16\pi^2}\,
  \lvert V_{tb} V_{ts}^* \rvert^2 \,
.\end{equation}
In $\bar B^0$ or $B^-$ [$B^0$ or $B^+$] decay, $\theta$ is the angle between the
$\ell^+$ [$\ell^-$] and the $B$ meson three-momenta in the $\ellpm$ center-of-mass
frame.  The Wilson coefficients $C_{7,9,10}$ contain short distance
information.  Beyond tree level they are effectively $q^2$ dependent and
complex, and receive different contributions in inclusive and exclusive decays,
as will be discussed below. If there were very precise data on $B\to X_s\ellpm$,
one could extract the individual Wilson coefficients from the $q^2$ dependence
of $\d\Gamma/\d q^2$ and the zero of the forward-backward asymmetry.  As long as
the measurements are limited by experimental uncertainties, it is important to
find the most effective ways to extract the short distance information from a
few simple observables integrated over $q^2$, accessible with a limited
amount of data.

In Section~\ref{sec:general} we discuss general aspects of an angular
decomposition, which gives three observables, $H_{T,A,L}(q^2)$. To extract short
distance information from these, we separate out the part of the Wilson
coefficients sensitive to new physics in a $q^2$ and $\mu$ independent manner,
and propose to compare measurements of these with their SM predictions.
Section~\ref{sec:inclusive} investigates inclusive $B\to X_s\ellpm$ decay,
while Section~\ref{sec:exclusive} deals with the exclusive $B\to K^*\ellpm$
mode. Section~\ref{sec:conclusions} contains our conclusions. Many analytical
results and numerical inputs are collected in the Appendices.

\section{Angular Decomposition and Dependence on Wilson Coefficients}
\label{sec:general}

The double differential decay rate in $q^2$ and $z = \cos\theta$ for either
inclusive $B\to X_s\ellpm$ or exclusive $B\to K^*\ellpm$ decays can be written
as
\begin{align}\label{doublediff}
\frac{\d^2\Gamma}{\d q^2\, \d z}
&= \frac{3}{8} \big[(1 + z^2) H_T(q^2) +  2 z H_A(q^2)
\nn\\
& \qquad + 2(1 - z^2) H_L(q^2) \big]
\,.\end{align}
The functions $H_i(q^2)$ are defined to contain the full $q^2$ dependence of the
rate and are independent of $z$. They can be extracted from the $\d^2\Gamma/\d
q^2\d z$ distribution either by a direct fit to the $z$ dependence or by taking
integrals of $z$. As special cases we have
\begin{align}
\frac{\d\Gamma}{\d q^2} &= H_T(q^2) + H_L(q^2)
\,,\nn\\
\frac{\d \AFB}{\d q^2} &= \frac{3}{4} H_A(q^2)
\,.\end{align}

For $H_L$ the hadronic current is longitudinally polarized, so the rate goes
like $\sin^2\theta = 1 - z^2$. For $H_T$ and $H_A$ the hadronic current is
transversely polarized, with $H_T$ containing the contributions from purely
vector and axial-vector leptonic currents and $H_A$ containing the interference
between vector and axial-vector leptonic currents. In the combinations $H_T \pm
H_A$ the hadronic and leptonic currents have the same (opposite) helicity,
giving the usual $(1 \pm \cos\theta)^2 = (1 \pm z)^2$ dependence.  This
decomposition is a common tool in the analysis of exclusive semileptonic decays
(e.g., $B\to D^{*}\ell\nu$, $\rho\ell\nu$, $K^{*}\ellpm$) and of decays to two
vector mesons (e.g., $B\to \phi K^*$, $J/\psi K^*$).  In analyzing
inclusive $B\to X_s\ellpm$, it should not be harder than measuring $A_{\rm FB}$.

We introduce a scheme to separate certain SM contributions to the rate from
terms that are most sensitive to new physics. We define modified Wilson
coefficients
\begin{align}\label{Ci_intro}
\C_7 &= C_7(\mu)\, \bigl[\overline m_b(\mu)/m_b^{1S}\bigr] + \ldots \,,\nn\\
\C_9 &= C_9(\mu) + \ldots \,,\nn\\
\C_{10} &\equiv C_{10} \,.
\end{align}
The ellipses denote a minimal set of perturbative corrections, such that
$\C_{7,9}$ are $\mu$ independent and real in the SM. These coefficients
are given explicitly in Eq.~\eqref{curlyC} in
Appendix~\ref{app:analytic} to $\ord(\alpha_s)$. The decay rate also depends on SM
contributions that are not contained in $\C_{7,9,10}$. These are discussed
in Secs.~\ref{sec:inclusive} and \ref{sec:exclusive}.  Of these, the dominant contributions are from the
four-quark operators $O_{1,2}$, which are expected to be given by the SM. We
regard the $\C_7$, $\C_9$, and $\C_{10}$ as the unknown parameters that need to
be extracted from experimental data, and compared with the SM or new physics
predictions.

We do not expand the $\overline{\rm MS}$ $b$-quark mass, $\overline{m}_b(\mu)$,
in $\C_7$, because it always comes together with $O_7$ and they should be
renormalized together. This also makes the perturbative expansions better
behaved. In addition, as indicated in Eq.~\eqref{Ci_intro}, we use the $1S$
scheme~\cite{Hoang:1998ng} for all other factors of $m_b$ (and $m_c$ as well),
which also improves the perturbative expansions. We will drop the superscript
$1S$ hereafter when the distinction is unimportant, but use $m_b^{1S}$ for $m_b$
everywhere (except $\overline{m}_b$, of course).

At subleading orders in $\alpha_s$ and $1/m_b$, the dependence of the rates on
the $\C_i$'s will be different in inclusive and exclusive decays.  To simplify
the explanation of our main points, in the remainder of this section we neglect
the contributions from operators other than $O_{7,9,10}$. Then, at leading
order, the $H_i$'s defined in Eq.~\eqref{doublediff} have the general structure
\begin{align}\label{Hi_general}
H_T(q^2) &\propto 2(1 - s)^2 s \Big[ \Big( \C_9 +
  \frac{2}{s}\, \C_7 \Big)^2 + \C_{10}^2 \Big]
,\nn\\
H_A(q^2) &\propto - 4(1 - s)^2 s\, \C_{10} \Big(\C_9 +
  \frac{2}{s}\, \C_7 \Big)
,\nn\\
H_L(q^2) &\propto (1 - s)^2 \big[ (\C_9 + 2 \C_7)^2 + \C_{10}^2 \big]
\,.\end{align}
Comparing Eqs.~\eqref{dGdq2} and \eqref{Hi_general} shows that splitting
$\d\Gamma/\d q^2$ into $H_L(q^2)$ and $H_T(q^2)$ separates the contributions
with different $q^2$ dependences, providing a third independent observable,
which has not been studied so far in inclusive $B\to X_s\ellpm$. For exclusive
$B\to K^*\ellpm$ it is equivalent to the fraction of longitudinal polarization
$F_L = H_L/(H_T + H_L)$ measured by Babar~\cite{Aubert:2006vb}.

If one does not resolve the details of the hadronic system, neglects effects
proportional to $m_\ell^2/m_b^2$, and does not measure the lepton polarization
(which might be accessible in $B\to X_s\tau^+\tau^-$~\cite{Hewett:1995dk},
though this is challenging at best), then the only observable linear combinations of the
Wilson coefficients are those given in Eq.~\eqref{Hi_general}. This is
necessarily the case for inclusive $B\to X_s e^+e^-$ and $B\to X_s\mu^+\mu^-$
decays.

In exclusive $B\to K^*\ellpm$ decay, if the $K^*\to K\pi$ decay is
reconstructed, there are two additional observable angles. These are
$\theta_{K^*}$, which is the analog of $\theta$ for the $K\pi$ system, and
$\phi$, which is the angle between the $K\pi$ and the $\ellpm$ planes (using the
notation of Babar~\cite{Aubert:2006vb}, which follows
Ref.~\cite{Kruger:2005ep}). Once one integrates over $\phi$, even if the
$\theta$ and $\theta_{K^*}$ distributions are not integrated over, the rate
depends only on the three linear combinations in Eq.~\eqref{Hi_general}. Keeping
the $\phi$ dependence would give rise to two further linear
combinations~\cite{Kruger:2005ep}, and it would require a more detailed study to
test whether the measurement could benefit from not integrating over $\phi$.

As mentioned above, instead of relying on the full $q^2$ dependence, we want to
integrate over as large regions of  $q^2$ as possible to extract the Wilson
coefficients from the simple integrals
\begin{equation}
H_i(q_1^2,\, q_2^2) = \int_{q_1^2}^{q_2^2} \d q^2\, H_i(q^2) \,
.\end{equation}
We restrict our discussion to the low $q^2$ region, $1\GeV^2 < q_1^2, q_2^2 <
6\GeV^2$, since it is theoretically clean and contains a large part of the rate.
The interference of the $J/\psi$ contribution with the short distance rate is a
significant contamination at higher values of $q^2$, while the rate for $q^2 >
m_{\psi'}^2 \approx 14.2\GeV^2$ is significantly smaller. Ultimately, the
measured tail of the long distance contribution will determine the optimal upper
cut on $q^2$.

For $H_L(q^2)$ the hadronic current is longitudinally polarized, so the $\C_7$
contribution is not enhanced by a $1/s$ pole, and is numerically small. Since
the Wilson coefficients in $H_L$ combine into a $q^2$ independent overall
factor, there is no gain in considering the $q^2$ dependence of $H_L(q^2)$.
Thus, to get maximal statistics one should use
\begin{equation}
H_L(1, 6) \propto \C_9^2 + \C_{10}^2 + 4\,\C_7 \C_9 + 4\,\C_7^2 \,,
\end{equation}
which is dominated by the $\C_9^2 + \C_{10}^2$ term.

In contrast, in $H_T(q^2)$ the different contributions have a
hierarchical $q^2$ dependence, i.e.
\begin{equation}
H_T(q^2) \propto \frac{4}{s}\, \C_7^2 + 4\,\C_7\C_9
  + s\, (\C_9^2 + \C_{10}^2) \,.
\end{equation}
In this case, integrating $H_T(q^2)$ over $q^2$, the $\C_7\C_9$ interference term in
$H_T(1,\,6)$ is as important as the $\C_9^2 + \C_{10}^2$ contribution, because
the latter vanishes for $q^2 \to 0$.  Thus, with enough statistics, it will be
worth splitting $H_T(1,6)$ into two integrals, $H_T(1,q_i^2)$ and $H_T(q_i^2,
6)$, giving access to two independent combinations of Wilson coefficients. The
value of $q_i^2$ should be chosen to give comparable statistics for the two
integrated rates. We will use $H_T(1,3.5)$ and $H_T(3.5,6)$ below.

Finally, $H_A(1,6) \propto \AFB$ has comparable $\C_7\C_{10}$ and $\C_9\C_{10}$
contributions. To separate them, one can split $H_A(1,6)$ into $H_A(1,q_i^2)$
and $H_A(q_i^2,6)$ similarly to $H_T$ above, where the precise value of $q_i^2$
can again vary. We will use $H_A(1,3.5)$ and $H_A(3.5,6)$, which give reasonably
independent linear combinations of $\C_7\C_{10}$ and $\C_9\C_{10}$.  We do not
normalize $H_A(q^2)$ by $\d\Gamma/\d q^2$, as is often done for $\d\AFB/\d q^2$.
This would suppress the low $q^2$ region, without providing any real
advantage. If necessary, $H_A(q^2)$ could be normalized to the rate in a certain
$q^2$ window.

Usually the zero of $H_A(q^2) \propto \d \AFB/\d q^2$, which occurs in the
vicinity of $q_{\rm FB}^2 = -2m_b^2(\C_7/\C_9)$, is advocated to determine
$\C_7/\C_9$.  As discussed in the introduction, measuring $q_{\rm FB}^2$
requires very large data sets.  Measuring two integrals of $H_A(q^2)$ as
described above may be a simpler way to achieve similar sensitivity with less
data.

\section{Inclusive {\boldmath $B\to X_{\!\lowercase{s}}\ellpm$}}
\label{sec:inclusive}

In this section we consider the inclusive decay $B\to X_s\ellpm$, working to
what is usually referred to as next-to-next-to-leading order (NNLO).  We define
the effective coefficients
\pagebreak
\begin{align}\label{Cincl}
\Cs(q^2) &= \C_7 + F_7(q^2) + G_7(q^2)
\,,\nn\\*
\Cn(q^2) &= \C_9 + F_9(q^2) + G_9(q^2)
\,,\end{align}
such that all terms on the right-hand side of Eq.~\eqref{Cincl} are separately
$\mu$ independent to the order we are working at.

The functions $F_i(q^2)$ and $G_i(q^2)$ are calculated in the SM. The
$F_{7,9}(q^2)$ contain perturbative contributions from the four-quark operators
$O_{1-6}$ and the chromomagnetic penguin operator, $O_8$, while the
$G_{7,9}(q^2)$ contain nonperturbative $\ord(1/m_c^2)$ corrections involving the
four-quark operators~\cite{Buchalla:1997ky}. The latter can be included in a
simple form for any differential rate, but the final results have to be
reexpanded so that $\ord(\alpha_s/m_c^2,\, 1/m_c^4)$ terms are not kept. The
explicit expressions are given in Appendix~\ref{app:analytic}. In the small
$q^2$ region (well below the $c\bar c$ threshold), the $\Csn(q^2)$ have only
small imaginary parts and modest $q^2$ dependences, which arise only from
$O_{1-6,8}$ and are fully contained in $F_i(q^2)$ and $G_i(q^2)$. Therefore,
all $\C_i$ are real numbers in the SM, which has the advantage of reducing the
number of parameters to three.

At NNLO, we include the corrections to the $F_i(q^2)$ up to $\ord(\alpha_s)$
where they are known analytically~\cite{Misiak:1992bc, Asatryan:2001de,
Beneke:2001at, Ghinculov:2003qd}, and to the Wilson coefficients of $O_{1-6,8}$
entering the $F_i(q^2)$~\cite{Adel:1993ah, Chetyrkin:1996vx, Bobeth:1999mk,
Bobeth:2003at}. At leading order in $\alpha_s$, $F_7(q^2)$ vanishes, while
$F_9(q^2)$ is nonvanishing. We also include the $\ord(\alpha_s)$ corrections to
the matrix elements of $O_{7,9,10}$~\cite{Asatryan:2001de, Ghinculov:2002pe,
Asatrian:2002va} and the nonperturbative $\ord(1/m_b^2)$~\cite{Falk:1993dh,
Ali:1996bm} corrections. The short distance contributions to the $H_i(q^2)$ can
be written as [$s = q^2/(m_b^{1S})^2$]
\begin{widetext}
\begin{align}
\label{Hincl}
H_T(q^2) &= 2\Gamma_0 (m_b^{1S})^3\, (1 - s)^2 s
\Big[(\Cns + \C_{10}^2) h_T^{99}(s)
  + \frac{4}{s^2} \Css h_T^{77}(s)
  + \frac{4}{s} \ReCsCn h_T^{79}(s) \Big]
  + H_T^\mathrm{brems}(q^2)
\,,\nn\\
H_A(q^2) &= - 4\Gamma_0 (m_b^{1S})^3\, (1 - s)^2 s\, \C_{10} \Re\Big[ \Cn h_A^{90}(s)
  + \frac{2}{s} \Cs h_A^{70}(s) \Big]
  + H_A^\mathrm{brems}(q^2)
\,,\\
H_L(q^2) &= \Gamma_0 (m_b^{1S})^3\, (1 - s)^2
  \Big[ \big(\Cns + \C_{10}^2\big) h_L^{99}(s) + 4 \Css h_L^{77}(s) +
  4 \ReCsCn h_L^{79}(s) \Big] + H_L^\mathrm{brems}(q^2)
\,. \nn
\end{align}
The functions $h_i^j(s)$ are defined to have only a residual $s$ dependence
entering at higher orders in $\alpha_s$ and $1/m_b$, i.e.
\begin{equation}
\label{hij_decomp}
h_i^j(s) = 1 - 2 \frac{\alpha_s C_F}{4\pi}\, \omega_i^j(s)
  + \frac{1}{m_b^2}\, \chi^j_i(s)
\,,\end{equation}
with $C_F = 4/3$.
The $\omega_i^j(s)$ and $\chi_i^j(s)$ containing the $\ord(\alpha_s)$
and $\ord(1/m_b^2)$ corrections are given in Appendix~\ref{app:analytic}.
We neglect the finite bremsstrahlung corrections due to four-quark operators,
$H_i^{\rm brems}(q^2)$, which are not known for the double differential rate
$\d^2\Gamma/\d q^2\d z$. This should be a safe approximation, since in the known
cases of $\d\Gamma/\d q^2$~\cite{Asatryan:2002iy} and $\d\AFB/\d
q^2$~\cite{Asatrian:2003yk} they are at or below the one percent level.

\begin{figure*}[t]
\centerline{\includegraphics[width=.45\textwidth]{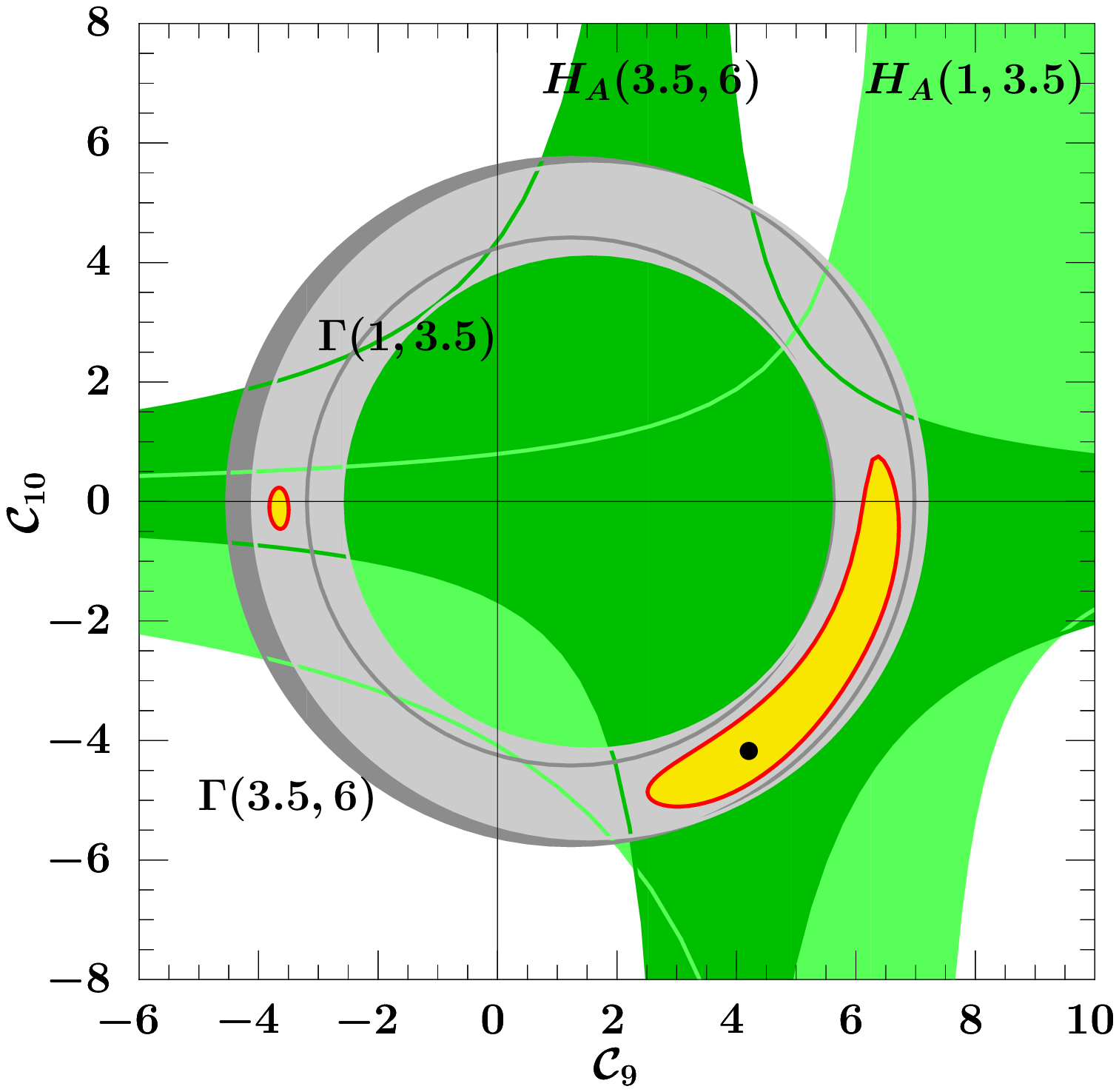}
\hspace{\columnsep}
\includegraphics[width=.45\textwidth]{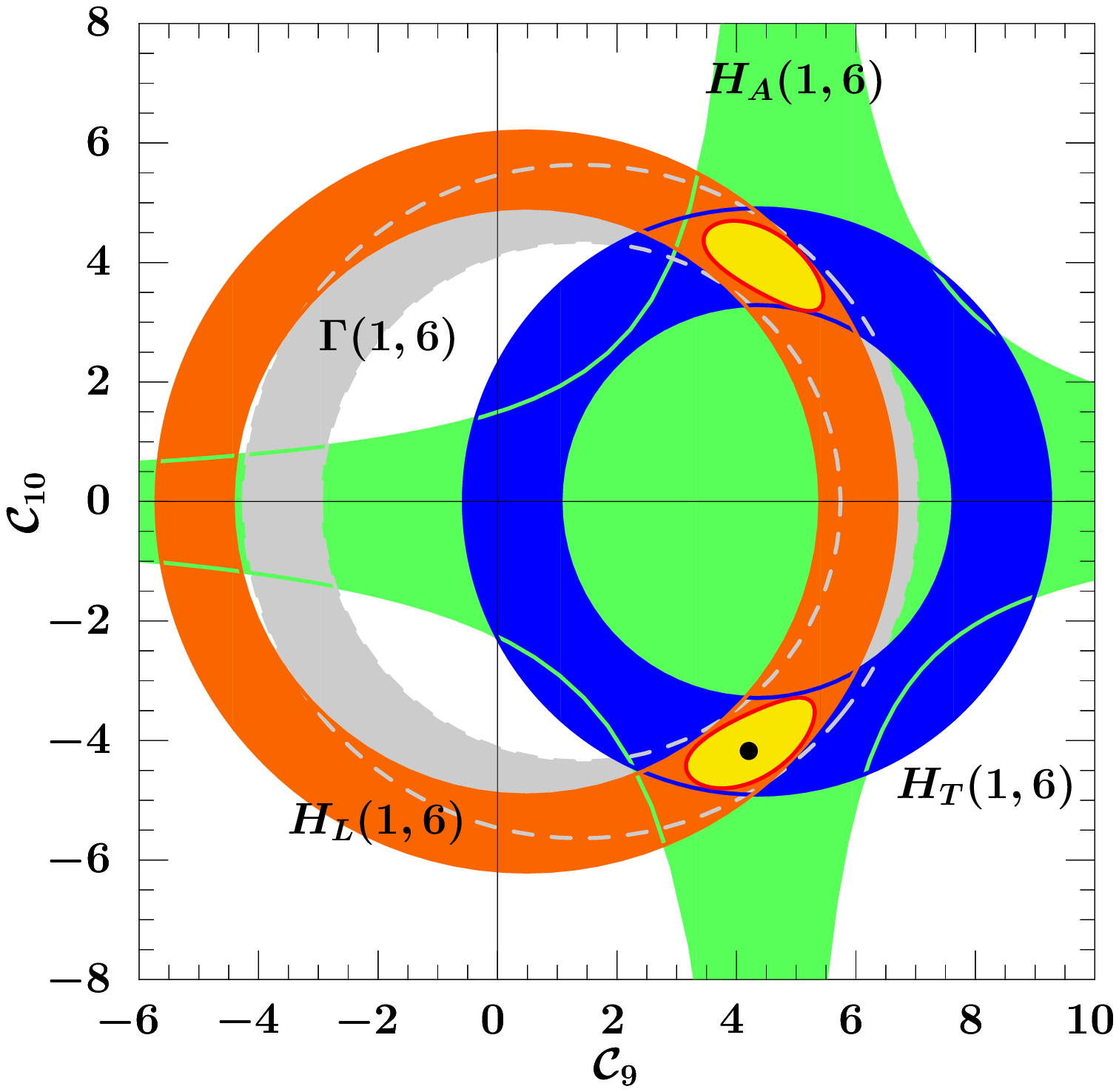}}
\caption{(color online) Constraints in the $\C_9-\C_{10}$ plane. Left:
$\Gamma(1,3.5)$ and $\Gamma(3.5,6)$ [light and medium (gray) annuli],
$H_A(1,3.5)$ and $H_A(3.5,6)$ [light and dark (green) regions bounded by
hyperbolae]. Right:  $H_T(1,6)$ [dark (blue) annulus],  $H_L(1,6)$ [medium
(orange) annulus], $H_A(1,6)$ [light (green) region bounded by hyperbolae], and
for comparison $\Gamma(1,6)$ [light (gray) annulus]. The black dots show the
assumed (SM) central values. The combined constraints are also shown [small
light (yellow) regions].}
\label{fig:C9C10}
\end{figure*}

To make numerical predictions we use the input values collected in
Appendix~\ref{app:numeric}. Since the value of $m_b^{1S}$ is known to better
than one percent, there would not be a significant benefit relative to other
uncertainties in normalizing the rates to $\Gamma(B\to X\ell\bar\nu)$. We find
\begin{align}\label{Hi_results}
H_T(1,3.5) / \Gamma_0  &=
   37.18\, (\C_9^2 + \C_{10}^2)
 + 15550.\, \C_7^2
 + 1409.\, \C_7 \C_9
 - 119.4\, \C_9
 - 2795.\, \C_7
 - 69.65
\,,\nn\\*
H_T(3.5,6) / \Gamma_0  &=
   59.76\, (\C_9^2 + \C_{10}^2)
 + 1067.\, \C_7 \C_9
 + 5008.\, \C_7^2
 - 76.25\, \C_9
 - 778.2\, \C_7
 - 67.72
\,,\nn\\
H_A(1,3.5) / \Gamma_0  &= - \C_{10}\, (
   70.19\, \C_9
 + 1401.\, \C_7
 - 121.5)
\,,\nn\\
H_A(3.5,6) / \Gamma_0  &= - \C_{10}\, (
   111.8\, \C_9
 + 1051.\, \C_7
 - 80.91)
\,,\nn\\*
H_L(1,6) / \Gamma_0 &=
   315.2\, (\C_9^2 + \C_{10}^2)
 + 1377.\, \C_7^2
 + 1299.\, \C_7 \C_9
 + 33.41\, \C_9
 + 86.86\, \C_7
 - 72.87
\,.\end{align}
\end{widetext}
The major uncertainties in Eqs.~\eqref{Hi_results} arise from higher order
perturbative corrections, $m_b$ and $m_c$. Varying the renormalization scale
between $m_b/2$ and $2m_b$, we get less than 5\% uncertainty in the coefficients
of the dominant terms in Eq.~(\ref{Hi_results}). Since the difference $m_b -
m_c$ is known precisely~\cite{Bauer:2004ve}, we vary $m_b$ and $m_c$ in a
correlated manner, which gives a 1--5\% uncertainty.  The uncertainties from
other input parameters and higher order corrections in $1/m_b$ are much
smaller. The uncertainties from the electroweak matching scale, $\mu_0$,
and the top-quark mass, $m_t$, in Eq.~\eqref{Hi_results} are negligible, because
they primarily enter via the values of the $\C_i$. Using  Eq.~\eqref{Hi_results}
and the values of $\C_i$ from Appendix~\ref{app:numeric}, we obtain the SM
prediction for the $B\to X_s\ellpm$ branching ratio for $1\GeV^2 < q^2 <
6\GeV^2$,
\begin{align}
\tau_B \Gamma(1,6)
&= \big(1.575 \pm 0.067_{[\mu]} \pm 0.051_{[m_b,m_c]}
\nn\\
& \quad \pm 0.041_{[m_t]} \pm 0.019_{[\mu_0]} \big) \times 10^{-6} \,.
\end{align}
This agrees well with Refs.~\cite{Ghinculov:2003qd,Bobeth:2003at} and
with Ref.~\cite{Huber:2005ig}, which also uses the $1S$ scheme.

\begin{figure*}[t]
\centerline{\includegraphics[width=.45\textwidth]{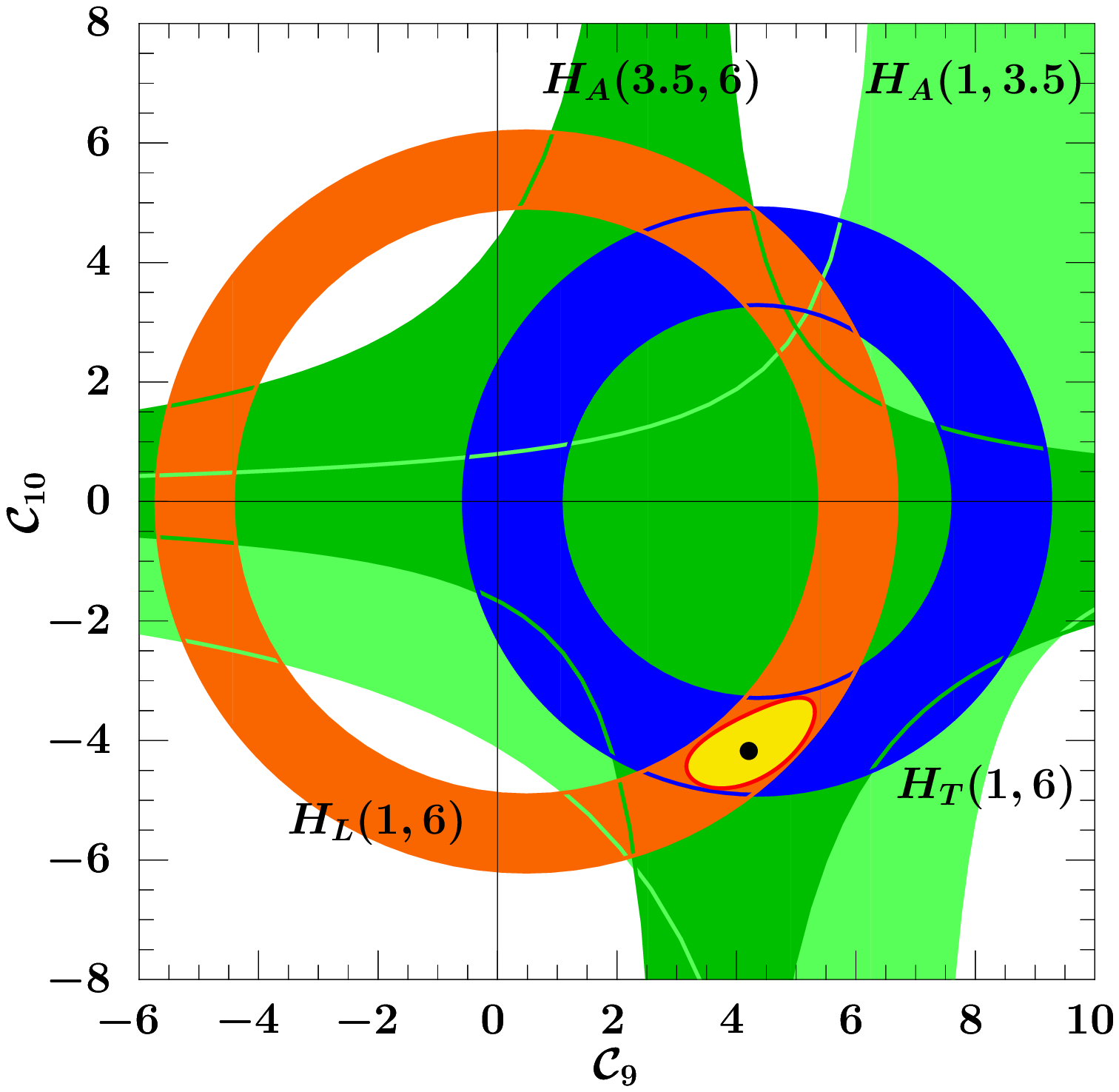}
\hspace{\columnsep}
\includegraphics[width=.45\textwidth]{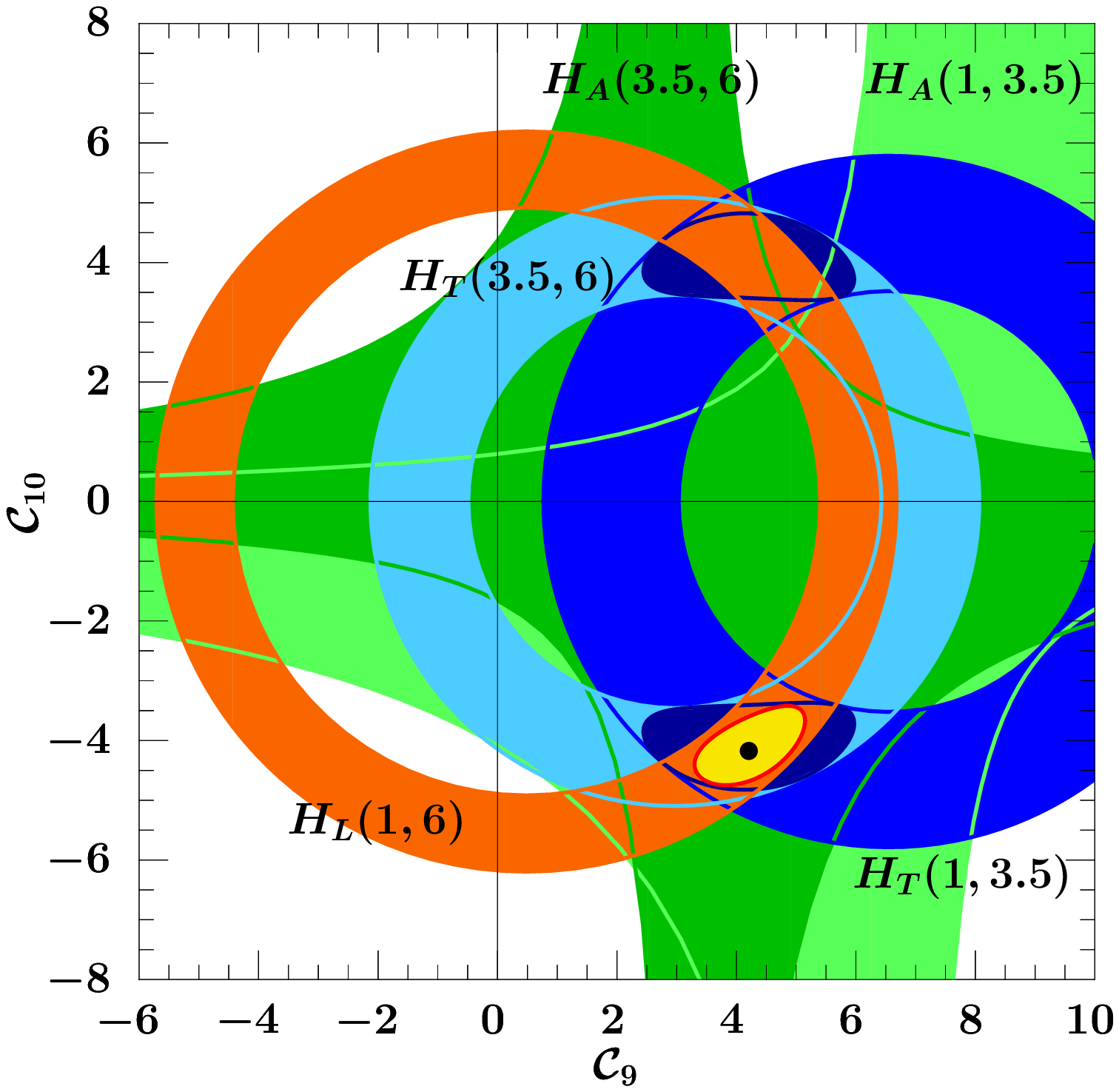}}
\caption{(color online) Constraints in the $\C_9-\C_{10}$ plane. Left: $H_T(1,
6)$ [dark (blue) annulus], $H_L(1,6)$ [medium (orange) annulus], $H_A(1,3.5)$
and $H_A(3.5,6)$ [light and dark (green) regions bounded by hyperbolae]. Right:
as on the left, but $H_T$ split into $H_T(1,3.5)$ and $H_T(3.5,6)$ [dark and
light (blue) annuli]. The right plot also shows the constraint from the two
$H_T$ observables alone [small black (dark blue) regions]. All other notations
are as in Fig~\ref{fig:C9C10}.}
\label{fig:C9C10_2}
\end{figure*}

To illustrate the improvement in determining the Wilson coefficients one
can obtain by separating $\Gamma$ into $H_T$ and $H_L$, we scale the current
measurements~\cite{Aubert:2004it,Iwasaki:2005sy} to $1\, \mathrm{ab}^{-1}$
luminosity, which gives about $10\%$ statistical uncertainty for $\Gamma(1,6)$.
We assume that the $H_i$ are measured with the central values given by the SM.
The statistical error of $H_T$ and $H_L$ is obtained by scaling by the number of
events compared to $\Gamma(1,6)$. In the case of $H_A$ we take the same absolute
statistical error for $3/4 H_A$ as for the total rate integrated over the same
$q^2$-region. The reason is that $3/4 H_A$ corresponds to the difference between
the rates for positive and negative $\cos\theta$, which has the same absolute
statistical error as the sum. To this we add in quadrature a $20\%$ systematic
uncertainty for all $H_i$, to account for experimental systematics and
theoretical uncertainties. From each observable's total error we build $\chi^2$
for the individual and combined constraints, and Figs.~\ref{fig:C9C10} and
\ref{fig:C9C10_2} show the $\Delta\chi^2 = 1$ regions in the $\C_9-\C_{10}$
plane. Since $B\to X_s\gamma$ will always be measured with higher precision than
$B\to X_s\ellpm$, we consider the value of $\C_7^2$ to be known from $B\to
X_s\gamma$ and assume its sign is negative as in the SM (since there is an
overall sign ambiguity).

On the left-hand side in Fig.~\ref{fig:C9C10} we show the constraints from
$\Gamma(1,3.5)$, $\Gamma(3.5,6)$ [light and medium (gray) annuli], $H_A(1,3.5)$,
and $H_A(3.5,6)$ [light and dark (green) regions bounded by hyperbolae]. This
plot shows that splitting $\Gamma(1,6)$ into two regions is not really useful
because a very similar linear combination of Wilson coefficients is
constrained.  As is well known, splitting $H_A(1,6)$ into two regions is very
useful, since different combinations of coefficients are constrained by each
region. (This is also the reason why the zero of the forward-backward asymmetry
is interesting to study.) The plot on the right in Fig.~\ref{fig:C9C10} shows
that splitting $\Gamma(1,6)$ into $H_T(1,6)$ [dark (blue) annulus] and
$H_L(1,6)$ [medium (orange) annulus] gives a very powerful constraint. This
shows the power of separately measuring $H_T$ and $H_L$ as advocated in the
introduction. The observables shown in this figure are sufficient to extract the
absolute values $\lvert\C_i\rvert$ and the sign of $\C_9$ relative to $\C_7$.
The constraint from $\Gamma(1,6)$ is also plotted [light (gray) annulus], which
shows that separating $\Gamma$ into $H_T$ and $H_L$ significantly improves the
constraints. However, because of its large relative error, $H_A(1,6)$ [light
(green) region bounded by hyperbolae] does not provide a good constraint.

The left plot in Fig.~\ref{fig:C9C10_2} shows that splitting $H_A$ into two
regions gives sensitivity to the sign of $\C_{10}$. Measuring $H_A(1,3.5)$ and
$H_A(3.5,6)$ can distinguish between the positive and negative solutions for
$\C_{10}$.  Combined with the tighter constraints from splitting $\Gamma$ into
$H_T$ and $H_L$, splitting $H_A$ into these two regions gives information similar
to the zero of $\AFB$.  The plot on the right in
Fig.~\ref{fig:C9C10_2} shows that splitting $H_T(1,6)$ into $H_T(1,3.5)$ and
$H_T(3.5,6)$ can further overconstrain the determination of the Wilson
coefficients. The black (dark blue) region in the right plot in
Fig.~\ref{fig:C9C10_2} shows the combined constraint from only the two $H_T$
integrals. The requirement that the $H_L$ constraint overlaps with it
effectively provides a consistency test on the value of $\C_7$ extracted from
$B\to X_s\gamma$.  Such overconstraining determinations of the Wilson
coefficients provide model independent searches for physics beyond the SM,
complementary to and possibly more effective than constraining specific new physics
scenarios (e.g., Wilson coefficients with flipped signs, complex values, or
adding right-handed operators).  For example, if there are operators with
right-handed helicity structure, they will affect $H_L$ and
$H_T$ differently, because of the different polarizations.

It would also be interesting to explore experimentally whether the influence of
the $J/\psi$ resonance turns on at similar $q^2$ values in $H_T$, $H_A$, and
$H_L$.  Since we have no information on the $J/\psi$ polarization in inclusive
$B\to J/\psi X_s$ decay, it is possible that the upper cut on $q^2$ can be
extended past $6\GeV^2$ in some (but maybe not all) of these observables,
which may improve the statistical accuracy of the measurement.

\section{Exclusive {\boldmath $B\to K^*\ellpm$}}
\label{sec:exclusive}

We now turn to the exclusive decay $B\to K^* \ellpm$.  While the theoretical
uncertainties are larger than in the inclusive analysis, measuring the exclusive
mode is simpler and it may be the only possibility at LHCb.  Compared with the
inclusive decay, the exclusive measurements go closer to $q^2 = m_{\psi}^2$:
$8.1\GeV^2$ at Belle~\cite{Abe:2004ir} and $8.4\GeV^2$ at
Babar~\cite{Aubert:2006vb}.  In this region of phase space the energy of the
$K^*$ varies only between $1.9\GeV < E_{K^*} < 2.7\GeV$, which helps control
some theoretical uncertainties.  In our general discussion we will consider for
simplicity $0.1\GeV^2 < q^2 < 8\GeV^2$, where the precise value of neither limit
is important (the lower limit can be replaced by any experimentally appropriate
value above $4m_\ell^2$).  However, for comparisons of our results with the
data, we use the limits used in the experimental analysis.

In this section we explain that, similarly to the inclusive decay, all the
information obtainable can be extracted from a few integrated rates.  To obtain
the most information on the ratios of Wilson coefficients,
Belle~\cite{Abe:2004ir} performed a maximum-likelihood fit to the double
differential distribution $\d^2\Gamma/\d q^2\,\d z$.  However, since the
theoretical predictions change for the double differential rate as they are
being refined, we think that a few integrated rates will be very useful to
compare the theory with the data. In addition, results from different
experiments are more straightforward to combine for these partial rates.

In the heavy quark limit, soft-collinear effective theory (SCET)~\cite{scet}
relates the seven full QCD form factors that describe $B\to K^* \ellpm$ to fewer
functions.  We follow the notation of Ref.~\cite{Bauer:2004tj}, and denote these
by $\zeta_{\perp,\parallel}$ and $\zeta^J_{\perp,\parallel}$.  The
$\zeta_{\perp,\parallel}$ obey the form factor relations~\cite{Charles:1998dr},
while $\zeta^J_{\perp,\parallel}$ violate them.  Whether $\zeta^J/\zeta$ is
${\cal O}(\alpha_s)$ or ${\cal O}(1)$ is subject to discussion, and so at the
present time the $\alpha_s$ corrections in exclusive decay are not fully known.

The angular decomposition for $B\to K^* \ellpm$
is~\cite{Ali:1999mm,Beneke:2001at,Pirjol:2003ef}
\begin{widetext}
\begin{align}\label{Hexcl}
H_T(q^2) &= 2\Gamma_0\, m_B^3 \lambda^3 s\,
  \bigg\{ \C_{10}^2\, [\zeta_\perp(s)]^2
  + \bigg| C_9^{\rm excl}\, \zeta_\perp(s) + \frac{2 C_7^{\rm excl}}{s}\,
  \frac{m_b^{1S}}{m_B}\,
  \big[\zeta_\perp(s) + (1-s)\, \zeta^J_\perp(s)\big] \bigg|^2 \bigg\}
\,,\nn\\
H_A(q^2) &= - 4\Gamma_0\, m_B^3 \lambda^3 s\; \C_{10}\, \zeta_\perp(s)
  \Re \bigg\{ C_9^{\rm excl}\, \zeta_\perp(s) + \frac{2C_7^{\rm excl}}{s}\,
  \frac{m_b^{1S}}{m_B}\,
  \big[ \zeta_\perp(s) + (1-s)\, \zeta^J_\perp(s) \big] \bigg\}
\,,\nn\\
H_L(q^2) &= \frac{1}{2}\, \Gamma_0\, m_B^3 \lambda^3\,
  \bigg( \C_{10}^2 + \bigg\lvert C_9^{\rm excl} + 2 C_7^{\rm excl}\,
  \frac{m_b^{1S}}{m_B} \bigg\rvert^2 \bigg)
  \big[\zeta_\parallel(s) - \zeta^J_\parallel(s)\big]^2
\,.\end{align}
\end{widetext}
In this section $s=q^2/m_B^2$.  In Eq.~(\ref{Hexcl}),  $\rho = m_{K^*}^2 / m_B^2
\sim 0.03$ and $\lambda = [(1-s)^2 - 2\rho\, (1+s) + \rho^2]^{1/2}$.
In analogy to Eq.~\eqref{Cincl}, we have defined
\begin{align}\label{Cexcl}
C_7^{\rm excl}(q^2) &= \C_7 + F_7(q^2) + \ord(\alpha_s)
\,,\nn\\
C_9^{\rm excl}(q^2) &= \C_9 + F_9(q^2) + \ord(\alpha_s)
\,.\end{align}
Note that there are additional ${\cal O}(\alpha_s)$ corrections beyond
$F_{7,9}(q^2)$~\cite{Beneke:2001at,Ali:2006ew}, and we do not attempt to address
power suppressed terms.

Without lattice QCD (LQCD) input, model calculations, or nonleptonic decay data to
constrain $\zeta_\parallel^{(J)}$, we cannot learn about the $\C_i$'s from
$H_L$.  However, the magnitudes of the form factors $\zeta_\perp^{(J)}$ can be
constrained from the $B\to K^*\gamma$ rate,
\begin{align}\label{BKsgamma}
\Gamma(B\to K^*\gamma) &= \frac{G_F^2}{8 \pi^3}\,
  \frac{\alpha_\mathrm{em}}{4\pi}\,
  \lvert V_{tb} V_{ts}^*\rvert^2\, m_B^3 (m_b^{1S})^2\, (1 - \rho)^3
\nn\\
& \quad \times \lvert C_7^{\rm excl}(0) \rvert^2\, \big[\zeta_\perp(0) + \zeta^J_\perp(0)\big]^2
\,.\end{align}
In  Eqs.~\eqref{BKsgamma}, (\ref{Hexcl}), and (\ref{Rdef}) we display the
kinematical $\rho$ dependences, but neglect the ``dynamical" ones, which would,
for example, multiply $\zeta^J_\perp(0)$ in Eq.~\eqref{BKsgamma} by $(1-\rho)$.

The $\zeta$ and $\zeta^J$ form factors are functions of the $K^*$ energy,
$E_{K^*} = (m_B/2)(1-s+\rho)$. In the heavy quark limit, a convenient
parameterization is
\begin{equation}\label{Edep}
\zeta_\perp^{(J)}(s) = \frac{\zeta_\perp^{(J)}(0)}{(1-s)^2}\,
  \big[1 + {\cal O}(\alpha_s, \Lambda/E_{K^*}) \big] \,,
\end{equation}
i.e., the form factors' leading $E$ dependence is $1/E_{K^*}^2$.  Since in the
$0.1\GeV^2 < q^2 < 8.4\GeV^2$ region $E_{K^*}$ varies only over $1.9\GeV <
E_{K^*} < 2.7\GeV$, we do not expect large deviations from this limit. For
example, $\zeta_\perp$ can have additional logarithmic dependence on
$E_{K^*}$~\cite{Lange:2003pk}, which is
roughly constant over this small region.

In the literature $\zeta^J/\zeta$ is often treated as ${\cal O}(\alpha_s)$.  The
zero-bin~\cite{Manohar:2006nz} shows that this is not the case parametrically,
so we treat both $\zeta_\perp(0)$ and $\zeta_\perp^J(0)$ as independent
nonperturbative parameters, and neglect $\ord(\alpha_s)$ corrections to
$\zeta_\perp^{(J)}(s)$, which are partially known.  These same considerations
also imply that the zero of the forward-backward asymmetry in $B\to K^*\ellpm$,
$q_{\rm FB}^2$, does not necessarily provide as precise and as model independent
a determination of $\C_7/\C_9$ as claimed in much of the literature.  Moreover,
even after 5 years of LHCb data taking (10\,fb$^{-1}$), one expects
$\sigma(q_{\rm FB}^2) \approx 0.5\GeV^2$, which would determine $\C_7/\C_9$ only
with an approximately $13\%$ error in the SM~\cite{LHCb}.

Some of the known ${\cal O}(\alpha_s)$ corrections to Eqs.~(\ref{Hexcl}) and
(\ref{BKsgamma}) can be included in Eq.~(\ref{Cexcl}).  However, since we treat
$\zeta_\perp(0)$ and $\zeta_\perp^J(0)$ as independent unknowns and determine
them from the data, only the part of the $\alpha_s$ corrections that causes these
form factors to deviate from their asymptotic $s$ dependences in
Eq.~(\ref{Edep}) will introduce errors.  Since in the $0.1\GeV^2 < q^2 < 8\GeV^2$
region that we concentrate on, $0 < s < 0.3$, we expect that the neglected
${\cal O}(\alpha_s)$ terms do not introduce a dominant error.  Moreover, they
could be added to our analysis.

From the ratios of the four observables
\begin{equation}\label{4excl}
\Gamma(B\to K^*\gamma), \quad H_T(0.1,8), \quad
  H_A(0.1,4), \quad H_A(4,8),
\end{equation}
one can extract
\begin{equation}\label{3solve}
\frac{\C_9}{\C_7}\,, \quad \frac{\C_{10}}{\C_7}\,, \quad
  r = \frac{\zeta_\perp^J(0)}{\zeta_\perp(0) + \zeta_\perp^J(0)}\,.
\end{equation}
If one does not want to use the $B\to K^*\gamma$ data then ratios of
$H_T(0.1,4)$, $H_T(4,8)$, $H_A(0.1,4)$, and $H_A(4,8)$ could also be used. The
ratio $r$ is of great interest for heavy quark theory.  One expects $r$ to be
roughly similar in size to the same ratio involving the $\zeta^{(J)}_\parallel$
form factors, which enters the determination of the unitarity triangle angle
$\gamma$ (or $\alpha$) from charmless two-body $B$ decays.

\begin{figure}[t]
\centering
\includegraphics[width=.95\columnwidth]{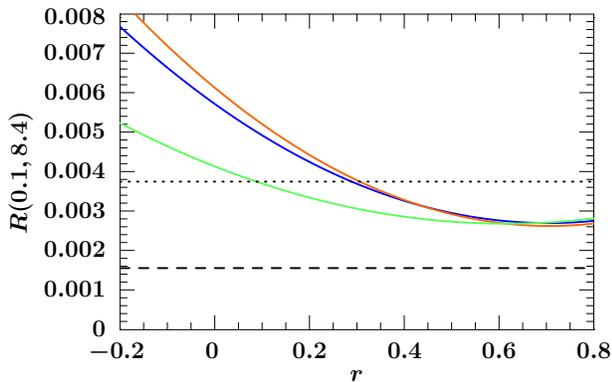}
\caption{(color online) The ratio $R(0.1, 8.4)$ defined in Eq.~\eqref{Rdef} as a function of
$r$ [dark (blue) curve], assuming the SM values for all $\C_i$. The medium (orange) curve
includes the $\ord(1)$ correction from $F_9(q^2)$, and the light (green) curve in addition
includes $\ord(\alpha_s)$ corrections from $F_{7,9}(q^2)$ in Eq.~\eqref{Cexcl}.
The dashed [dotted] line shows the central value [$1\sigma$ upper bound] from
the Babar measurement~\cite{Aubert:2006vb}.}
\label{fig:R}
\end{figure}

At present there is insufficient published data to carry out this analysis to
determine the Wilson coefficients.  However, $\Gamma(B\to K^*\gamma)$ is well
measured~\cite{pdg} and $H_T(0.1,8.4)$ can be obtained from the Babar
measurement~\cite{Aubert:2006vb}.  (Belle~\cite{Abe:2004ir} does a
maximum-likelihood fit to extract information on the Wilson coefficients, so we
cannot use their fit result projected on to the forward-backward asymmetry as a
direct measurement of $H_A$ in the corresponding bins.)   Assuming the SM for
$\C_{7,9,10}$, we can use $\Gamma(B\to K^*\gamma)$ and $H_T(0.1,8.4)$ to
constrain $r$. The ratio
\begin{align}\label{Rdef}
&R(q_1^2,q_2^2) \equiv \frac{H_T(q_1^2,q_2^2)}{\Gamma(B\to K^*\gamma)} \nn\\
&= \frac{\alpha_\mathrm{em}}{12\pi} \frac{m_B^2}{(m_b^{1S})^2}
  \int_{q_1^2/m_B^2}^{q_2^2/m_B^2} \d s\,
  \frac{\lambda^3\, s}{(1-\rho)^3\, (1 - s)^4} \\
&\times\! \biggl\{\! \frac{\C_{10}^2}{\C_7^2} (1-r)^2
  + \biggl[\frac{\C_9}{\C_7} (1-r)
  + \frac2{s} \frac{m_b^{1S}}{m_B} (1-s r) \biggr]^2 \!\!+ \dots\biggr\}, \nn
\end{align}
as a function of $r$ is shown in Fig.~\ref{fig:R} for $R(0.1,8.4)$.  The dark (blue)
curve shows Eq.~\eqref{Rdef}, the medium (orange) one includes the leading corrections to
$C_9^{\rm excl}(q^2)$ from $F_9(q^2)$, and the light (green) curve includes in addition
the $\ord(\alpha_s)$ corrections from $F_{7,9}(q^2)$. The significant change is
dominantly due to the large $F_7(0)$ contribution to $C_7^{\rm excl}(0)$, also
observed for $\Gamma(B\to K^*\gamma)$~\cite{Bosch:2001gv,Beneke:2001at}. This
shows that a complete understanding of the $\alpha_s\, \zeta_\perp^J$
corrections to these exclusive decays is very important. Until this is achieved,
a determination of $\C_9/\C_7$, $\C_{10}/\C_7$, and $r$ without the $\Gamma(B\to
K^*\gamma)$ data, using only two bins of each $H_T$ and $H_A$, as mentioned after
Eq.~\eqref{3solve}, may be theoretically cleaner.

The central value, $R(0.1,8.4) = 1.55 \times 10^{-3}$~\cite{Aubert:2006vb},
and the 1$\sigma$ upper bound are shown in Fig.~\ref{fig:R} by dashed and dotted
lines, respectively. Since the error is still large, at the moment one cannot
make a statistically significant statement about the size of $r$. Nevertheless,
we expect that (if the SM is valid) the central value of $R(0.1,8.4)$ should
go up, probably via an increase in the transverse polarization fraction in this
$q^2$ region.

Until precise unquenched LQCD calculations of the $B\to K^{(*)}$ form factors
for small $q^2$ become available, the method outlined above may provide the most
accurate extraction of short distance information from $B\to K^*\ellpm$.  With
more statistics in the future it should become possible to determine the
quantities in Eq.~\eqref{3solve}, providing new tests of the SM and insights
into the theory of hadronic $B$ decays.

We did not consider $B\to K \ellpm$ decay, because $B\to K\gamma$ is forbidden
by angular momentum conservation, so it is not possible to learn about the short
distance physics from this mode without using a determination of the
corresponding form factors from lattice QCD or model calculations. (This
is similar to the case of $H_L^{B\to K^*\ellpm}$ explained above.)  To proceed
by using model independent continuum methods, one would have to use $B\to
\pi\ell\bar\nu$ data combined with $SU(3)$ flavor symmetry or the two-body
charmless nonleptonic decay data to constrain the $B\to K$ form factor.

\section{Conclusions}
\label{sec:conclusions}

In this paper we pointed out that in inclusive $B\to X_s\ellpm$ decay an angular
decomposition provides a third $q^2$ dependent observable in addition to the
total rate and the forward-backward asymmetry.  Splitting up the rate into
transverse and longitudinal parts, proportional to $1+\cos^2\theta$ ($H_T$) and
$1-\cos^2\theta$ ($H_L$), one gains access to a third independent linear
combination of Wilson coefficients.  This requires measuring no additional
kinematical variable besides $q^2$ and $\cos\theta$, which are already studied
by Babar and Belle. Without doing more complicated analyses, it will improve the
determination of the relevant Wilson coefficients and the sensitivity to
possible non-SM physics.

To incorporate the existing NNLO calculations, we proposed a new scheme that
defines $q^2$ independent coefficients, $\C_7$ and $\C_9$, which are real in the
SM.  The $\C_{7,9}$ do not contain certain SM contributions (involving
$C_{1-6,8}$), which make the coefficients usually referred to in the literature
as $C_{7,9}^{\rm eff}$ complex and $q^2$ dependent.  We view $\C_7$, $\C_9$, and
$\C_{10} \equiv C_{10}$ as the unknowns sensitive to physics beyond the SM
to be extracted from data.

Since precise measurements of the $q^2$ dependences require very large data
sets, we studied how one can extract all Wilson coefficients obtainable from
$B\to X_s\ellpm$ from a few simple integrals of the $H_{T,L,A}$ components of
the decay rate. We concentrated on the low $q^2$ region ($1\GeV^2 < q^2 <
6\GeV^2$) and found (see Figs.~\ref{fig:C9C10} and~\ref{fig:C9C10_2}) that
splitting the total rate into transverse and longitudinal parts is a powerful
tool to gain more information.

The same angular decomposition in exclusive $B\to K^*\ellpm$ decay, together
with the well-measured $B\to K^*\gamma$ rate, also provides a determination of
the Wilson coefficients, without reliance on form factor models and without
requiring a measurement of the zero of the forward-backward asymmetry.  In the
heavy quark limit, as a starting point of a systematic expansion, one can
parameterize the form factors in the low $q^2$ region by just a few numbers [see
Eq.~\eqref{Edep}].  Measuring the observables in Eq.~\eqref{4excl}, one can
extract from the data both the hadronic unknowns and the Wilson coefficients. A
more complete understanding of the $B\to K^*\ellpm$ decay may be expected in the
near future and will help to firm up the error estimates in such an analysis
based on the $B\to K^*\ellpm$ data.

It is not known if a truly inclusive study of $B\to X_d\ellpm$ will ever be
feasible experimentally, but it may be possible to study the exclusive decay
$B\to \rho\, \ellpm$.  The methods discussed in this paper are clearly
applicable to these decays as well.

With significantly more data, one may prefer to split the rate into more than
two bins in the $1\GeV^2 < q^2 < 6\GeV^2$ region.  One can then search for
non-SM physics by fitting for complex $\C_{7,9,10}$ values, or allowing opposite
chirality operators absent in the SM (model independently there is no preference
which way to extend the parameter space).  Given the good overall consistency of
the SM, overconstraining determinations of the Wilson coefficients, such as that
in Fig.~\ref{fig:C9C10_2}\,b, may give the best sensitivity to new physics
(similarly to the CKM fit).

We did not include shape function effects~\cite{Lee:2005pw} in our analysis for
the inclusive decay (nor are they included in any other paper performing fits to
extract short distance physics from the low $q^2$ region). This is left for
future work.  Based on Ref.~\cite{Lee:2005pw}, we anticipate that if the $B\to
X_s\gamma$ photon spectrum is used to understand the effect of the $m_X$ cut in
$B\to X_s\ellpm$, then the analysis considered in this paper will receive only
modest corrections, leaving the general picture of how best to extract the
Wilson coefficients unchanged.

\acknowledgments

We are grateful to Peter Cooper, Enrico Lunghi and especially Jeff Berryhill for
helpful conversations.
We thank Christoph Bobeth for providing us with his Mathematica code for the
running of the Wilson coefficients as given in Refs.~\cite{Bobeth:1999mk,
Bobeth:2003at}.
Z.L.\ thanks the MIT CTP for its hospitality while part of this work
was completed.
This work was supported in part by the Director, Office of Science, Offices
of High Energy and Nuclear Physics of the U.S.\ Department of Energy under the
Contracts DE-AC02-05CH11231 (Z.L.\ and F.T.), and the cooperative research
agreement DOE-FC02-94ER40818 (I.S.).  I.S. was also supported in part by the DOE
OJI program and by the Sloan Foundation.

\appendix
\section{Analytical Results}
\label{app:analytic}

Here we collect the explicit results for all $\alpha_s$ and power corrections
used in the main text. The Wilson coefficients $C_i(\mu)$ refer to the operator
basis of Refs.~\cite{Chetyrkin:1996vx, Bobeth:1999mk}, except that we keep
$C_{7-10}$ in the traditional normalization~\cite{Buchalla:1995vs}. Their
numerical values are collected in Table~\ref{Ci_numerical} below. (The
corresponding coefficients in the normalization of Ref.~\cite{Bobeth:1999mk} are
$\alpha_s/(4\pi) C_{7-10}$.) Since the formally leading term in $C_9(m_b)$ is
numerically small, it is often considered as $\ord(1)$ in the recent literature.
In our case this is not an issue because we treat $\C_9$ as an unknown $\ord(1)$
parameter to be extracted from experiment.

Throughout this paper we work in the $1S$ scheme~\cite{Hoang:1998ng}, with
$m_{c,b}$ always referring to the $1S$ masses $m_{c,b}^{1S}$. They
are related to the pole masses, $m^\mathrm{pole}$, by ($C_F = 4/3$)
\begin{equation}
\label{m1S_def}
m^{1S} = m^\mathrm{pole}
\bigg[1 - \frac{\alpha_s(\mu)\, C_F}{4\pi}\, \delta^{1S}(\mu)
+ \ord\big(\alpha_s^2 \delta^{1S} \big) \bigg]
\,,\end{equation}
where $\delta^{1S}(\mu) = (\pi/2) \alpha_s(\mu) C_F$ is formally counted as
$\ord(1)$. By definition $m^{1S}$ is $\mu$ independent. The $\mu$ dependence
reenters when the perturbative expansion on the right-hand side of
Eq.~\eqref{m1S_def} is truncated. Switching from the pole to the $1S$ scheme, one
reexpands the perturbative expressions to a given order in $\alpha_s$, counting
$\delta^{1S} \sim\ord(1)$, and using the same scale $\mu$ in Eq.~\eqref{m1S_def}
as everywhere else. All perturbative expressions below have been converted to
the $1S$ scheme. The expressions in the pole scheme given in the literature are
recovered by setting $\delta^{1S}(\mu) = 0$ and $m_b^{1S} = m_b^{\rm pole}$
everywhere.

In the following, repeated indices are summed from $1$ to $6$.
The coefficients $\C_{7-10}$ in Eq.~\eqref{Cincl} are defined as
\begin{widetext}
\begin{align}\label{curlyC}
\C_7 &= C_7(\mu) \frac{\overline{m}_b(\mu)}{m_b^{1S}}
  + C_i(\mu) \kappa_{i7} - \frac{\alpha_s(\mu)}{4\pi}\, \ln\frac{\mu}{m_b^{1S}}\,
  \bigg[ \frac{8}{3} C_7(\mu) \frac{\overline{m}_b(\mu)}{m_b^{1S}} - \frac{32}{9} C_8(\mu)
  + C_i(\mu) \Big(\gamma^{(0)}_{ij} \kappa_{j7} + \gamma_{i7}^{(0)} \Big) \bigg]
 + \ord(\alpha_s^2)
\,,\nn\\
\C_8 &= C_8(\mu) + C_i(\mu) \kappa_{i8} + \ord(\alpha_s)
\,,\nn\\
\C_9 &= C_9(\mu)
  + C_i(\mu) \bigg[ \kappa_{i9} - \gamma_{i9}^{(-1)} \ln\frac{\mu}{m_b^{1S}}
  - \frac{\alpha_s(\mu)}{4\pi}\, \ln\frac{\mu}{m_b^{1S}}\,
  \bigg( \gamma^{(0)}_{ij} \kappa_{j9} +  \gamma_{i9}^{(0)}
  - \frac{1}{2} \gamma_{ij}^{(0)} \gamma_{j9}^{(-1)} \ln\frac{\mu}{m_b^{1S}} \bigg) \bigg]
 + \ord(\alpha_s^2)
\,,\nn\\
\C_{10} &= C_{10}
\,,\end{align}
where the higher order $\alpha_s$ corrections are determined
by the requirements that they vanish at $\mu = m_b^{1S}$ and that the $\C_i$
are $\mu$ independent to a given order.
For convenience we also defined $\C_8$. It enters the $F_i(q^2)$ (see
below) at $\ord(\alpha_s)$, so we need it only at lowest order.
The relevant entries of the anomalous dimension matrices
are~\cite{Misiak:1992bc, Chetyrkin:1996vx, Bobeth:1999mk}
\begin{equation}
\gamma_{ij}^{(0)} =
\begin{pmatrix}
 -4 & \frac{8}{3} & 0 & -\frac{2}{9} & 0 & 0 \\
 12 & 0 & 0 & \frac{4}{3} & 0 & 0 \\
 0 & 0 & 0 & -\frac{52}{3} & 0 & 2 \\
 0 & 0 & -\frac{40}{9} & -\frac{100}{9} & \frac{4}{9} &
  \frac{5}{6} \\
 0 & 0 & 0 & -\frac{256}{3} & 0 & 20 \\
 0 & 0 & -\frac{256}{9} & \frac{56}{9} & \frac{40}{9} &
  -\frac{2}{3}
\end{pmatrix}
,\qquad
\begin{aligned}
\gamma_{i7}^{(0)} &= \bigg(-\frac{232}{243}\,,\frac{464}{81}\,,\frac{64}{81}\,,
-\frac{200}{243}\,,-\frac{6464}{81}\,,-\frac{11408}{243} \bigg)
\,,\\
\gamma_{i9}^{(-1)} &= \bigg(-\frac{32}{27}\,,-\frac{8}{9}\,,-\frac{16}{9}\,,
\frac{32}{27}\,, -\frac{112}{9}\,,\frac{512}{27} \bigg)
\,,\\
\gamma_{i9}^{(0)} &= \bigg(-\frac{2272}{729}\,,\frac{1952}{243}\,,-\frac{6752}{243}\,,
-\frac{2192}{729}\,,-\frac{84032}{243}\,,-\frac{37856}{729} \bigg)
\,.\end{aligned}
\end{equation}
The constants $\kappa_{ij}$ in Eqs.~\eqref{curlyC} are included by convention, as in
the usual definitions of $C_i^\mathrm{eff}$, and are given by~\cite{Chetyrkin:1996vx, Bobeth:1999mk}
\begin{equation}
\kappa_{i7} = \bigg(0\,, 0\,, -\frac{1}{3}\,, -\frac{4}{9}\,, -\frac{20}{3}\,, -\frac{80}{9} \bigg)
\,,\qquad
\kappa_{i8} = \bigg(0\,, 0\,, 1\,, -\frac{1}{6}\,, 20\,, -\frac{10}{3} \bigg)
\,,\qquad
\kappa_{i9} = \bigg(0\,, 0\,, 0\,, \frac{4}{3}\,, \frac{64}{9}\,, \frac{64}{27} \bigg)
\,.\end{equation}

The functions $F_i(q^2)$ in Eq.~\eqref{Cincl} contain perturbative
corrections arising from $O_{1-6,8}$. To $\ord(\alpha_s)$,
\begin{align}
\label{Fi_explicit}
F_7(q^2)
&= - \frac{\alpha_s(\mu)}{4\pi} \Big[ C_i(\mu)  f_i^{(7)}(\hat{m}_c,s) + \C_8 f_8^{(7)}(s) \Big]
,\nn\\
F_9(q^2)
&= C_i(\mu) \bigg[ \delta_{ij} - \frac{\alpha_s(\mu)}{4\pi}
\ln\frac{\mu}{m_b^{1S}} \gamma^{(0)}_{ij}
\bigg]
\big[ t_j h(\hat{m}_c, s) + u_j h(1, s) + w_j h(0, s) \big]
 - \frac{\alpha_s(\mu)}{4\pi}
 \Big[  C_i(\mu) f_i^{(9)}(\hat{m}_c,s) + \C_8 f_8^{(9)}(s) \Big]
\nn\\
& \quad - \frac{\alpha_s(\mu)}{4\pi} \frac{16}{9}\,\delta^{1S}(\mu) C_i(\mu)
\bigg[ t_i\frac{\mathcal{F}[s/(4\hat m_c^2)]}{1-s/(4\hat m_c^2)}
 + u_i\frac{\mathcal{F}(s/4)}{1-s/4} - \frac{2}{3} (t_i+u_i+w_i) \bigg]
,\end{align}
where $s = q^2/(m_b^{1S})^2$, $\hat{m}_c = m_c^{1S}/m_b^{1S}$, and
\begin{equation}
t_i = \bigg(\frac{4}{3}\,, 1\,, 6\,, 0\,, 60\,, 0 \bigg)
\,,\quad
u_i = \bigg(0\,, 0\,, -\frac{7}{2}\,, -\frac{2}{3}\,, -38\,, -\frac{32}{3} \bigg)
\,,\quad
w_i = \bigg(0\,, 0\,, -\frac{1}{2}\,, -\frac{2}{3}\,, -8\,, -\frac{32}{3} \bigg)
.\end{equation}
The second line in $F_9(q^2)$ arises from reexpanding the leading order
contribution in the $1S$ scheme. The one-loop function $h(\hat{m}_c,s)$ encoding
the four-quark contributions is~\cite{Grinstein:1988me,Misiak:1992bc}
\begin{align}
h(\hat{m}_c, s)
&= - \frac{8}{9} \ln(\hat{m}_c) - \frac{4}{27} + \frac{2s}{9\hat m_c^2}
  - \frac{4}{9} \Big(1+\frac{s}{2\hat m_c^2}\Big)\,\mathcal{F}\Big(\frac{s}{4\hat m_c^2}\Big)
\,,\nn\\
h(0,s) &= - \frac{4}{9} \ln(s) + \frac{8}{27} + \frac{4}{9}\, i \pi
\,,\end{align}
and we have defined the function
\begin{equation}
\mathcal{F}(x) = 1 - \frac{1}{x} + \frac{\sqrt{\lvert 1-1/x \rvert}}{x}
\begin{cases} \displaystyle
\arctan \sqrt{ \frac{x}{1-x}} , & 0<x<1 \,,\\[2ex]
\ln\big(\sqrt x + \sqrt{x-1}\, \big) - i\pi/2, \quad & x>1
\,.\end{cases}
\end{equation}
It satisfies $\mathcal{F}(0)=2/3$ and $\mathcal{F}(1-\epsilon) =
\sgn(\epsilon)(\pi/2)\sqrt{\epsilon} + \ord(\epsilon)$.

The one-loop functions $f_8^{(7,9)}(s)$ in Eq.~\eqref{Fi_explicit} containing
the $O_8$ contributions are
\begin{align}
f_8^{(7)}(s) &= -\frac{2}{9} \bigg[
\frac{4s}{1 - s}\ln(s) + (22 + 60 s +158 s^2 + 329 s^3 + \ldots)
  - (2 + 9s + 24s^2 + 50 s^3 + \ldots)\frac{2\pi^2}{3}  + 4i\pi \bigg]
,\nn\\
f_8^{(9)}(s) &= \frac{4}{9} \bigg[
  \frac{4}{1-s} \ln(s) + \frac{1}{15} (390 + 1480 s + 3553 s^2 + \ldots)
  - (4 + 15s + 36 s^2 + \ldots)\frac{2\pi^2}{3}  \bigg]
.\end{align}
\end{widetext}
Their exact analytic expressions in terms of integrals can be found in
Refs.~\cite{Beneke:2001at, Ghinculov:2003qd}. Here we expanded the terms not
proportional to $\ln(s)$ for small $s$. Since these are small compared with the
$\ln(s)$ term, the expressions above are accurate to better than $10^{-3}$ for
$s < 1$ and $10^{-5}$ for $s < 0.4$.

The two-loop functions $f_{1-6}^{(7,9)}$ in Eq.~\eqref{Fi_explicit} contain the
virtual $\ord(\alpha_s)$ contributions from $O_{1-6}$ and are known for $i = 1$
and $2$ only. In Ref.~\cite{Asatryan:2001de} they are given as expansions in $s$
and $\hat{m}_c$. Since the expressions for general $\hat{m}_c$ are quite
lengthy, Ref.~\cite{Asatryan:2001de} also quotes them at five fixed values for
$\hat{m}_c$, which we use together with a linear interpolation for intermediate
values of $\hat{m}_c$.

Usually, the contributions from $O_{1,2,8}$ that enter $\Csn(q^2)$ are included
via functions $F_{1,2,8}^{(7,9)}$~\cite{Asatryan:2001de}; e.g., for the $O_1$
contribution to $\Cn$
\begin{equation}
\Cn = C_9 - \frac{\alpha_s}{4\pi}\, C_1 F_1^{(9)} + \ldots\,
.\end{equation}
The $F_{1,2,8}^{(7,9)}$ consist of terms containing powers of $\ln(\mu/m_b)$ and
the functions $f_{1,2,8}^{(7,9)}$. For example, switching to the pole scheme,
the terms proportional to $\alpha_s/(4\pi) C_1$ are
\begin{align}
F_1^{(9)}
&= \ln\frac{\mu}{m_b} \bigg\{ \gamma^{(0)}_{1j} \kappa_{j9}
  + \gamma_{19}^{(1)}
  - \frac{1}{2} \gamma_{1j}^{(0)} \gamma_{j9}^{(0)} \ln\frac{\mu}{m_b} \nn\\
&\quad + \gamma^{(0)}_{1j} \big[ t_j h(\hat{m}_c, s) + u_j h(1,s) + w_j h(0,s)
  \big] \bigg\} + f_1^{(9)} \nn\\
&= -\ln\frac{\mu}{m_b}\bigg[ \frac{256}{243} \ln\frac{\mu}{m_b}
  - \frac{2272}{729} -\frac{8}{3} h(\hat{m}_c,s) \nn\\
&\quad + \frac{4}{27} h(1,s) + \frac{4}{27} h(0,s) \bigg] + f_1^{(9)}
.\end{align}
In our scheme these $\ln(\mu/m_b)$ dependent terms are split up between
$F_9(q^2)$ and $\C_i$, and are included in the definitions~\eqref{curlyC} and
\eqref{Fi_explicit}.
Expanding the term in brackets for small $s$, we recover the result given in
Ref.~\cite{Asatryan:2001de}. The same is true for all $F_{1,2,8}^{(7,9)}$. Note
that in this way we automatically include the $\ln(\mu/m_b)$ dependent terms of
the analogous functions $F_{3-6}^{(7,9)}$, which to our knowledge have not been
calculated explicitly so far.

The functions $G_{7,9}(q^2)$ contain $\ord(1/m_c^2)$ corrections in
Eq.~\eqref{Cincl}, calculated for $\d\Gamma/\d q^2$ and $\d\AFB/\d q^2$ in
Ref.~\cite{Buchalla:1997ky}. We found that their contribution  can be included
similarly to other four-quark operator contributions in
$C_{7,9}^\mathrm{incl}(q^2)$ [see Eq.~\eqref{Cincl}] via the functions
\begin{align}\label{1mc2}
G_9(q^2)
&= \frac{10}{1-2s}\, G_7(q^2)
\nn\\
&= - \frac{5}{6} \biggl(C_2 - \frac{C_1}6 \biggr) \frac{\lambda_2}{m_c^2}\,
  \frac{\mathcal{F}[q^2/(4m_c^2)]}{1-q^2/(4 m_c^2)}
\,.\end{align}
(In the operator basis used in Ref.~\cite{Buchalla:1997ky} the $C_2-C_1/6$ term
in Eq.~\eqref{1mc2} should be replaced by their $C_2$, and we set this
coefficient to unity in our numerical analysis.)
These corrections diverge as $(4m_c^2-q^2)^{-1/2}$ as $q^2$ approaches the
$4m_c^2$ threshold. Following Ref.~\cite{Buchalla:1997ky}, we imagine that this
calculation makes sense for $q^2 \lesssim 3m_c^2 \approx 6\GeV^2$.  Even for
$q^2 \ll 4m_c^2$, there is an infinite series of corrections suppressed only by
increasing powers of $\Lambda m_b/m_c^2$~\cite{Ligeti:1997tc}.

The $\omega_i^j(s)$ containing the $\ord(\alpha_s)$ corrections in
Eq.~\eqref{hij_decomp} can be extracted from Ref.~\cite{Asatrian:2002va}.
Defining
\begin{equation}
L(s) = 4 \Li_2(s) + 2 \ln(s)\ln(1-s) + \frac{2\pi^2}{3}
\,,\end{equation}
we find
\begin{widetext}
\begin{align}
\omega_T^{99}(s) &=
  \frac{1}{2\sqrt{s}}\, \omega^{99}(s) + L(s) + \frac{1+2s}{s} \ln(1-s)
+ \frac{5 + s(7-2s)}{(1-s)^2} \ln(s)
+ \frac{19+s}{2(1-s)}
- \frac{1 + 3s}{1 - s} \,\frac{\delta^{1S}(\mu)}{2}
\,,\nn\\
\omega_A^{90}(s) &=
 2\frac{1 + s(3-s)}{(1-s)^2} \Li_2(s)
+ \frac{1 - s(9-2s)}{s(1-s)} \ln(1-s)
- \frac{4s(5-2s)}{(1-s)^2} \Li_2(\sqrt{s})
+ \frac{2(5-2s)}{1-s} \ln(1-\sqrt{s}) \nn\\
& \quad + \frac{2+s}{(1-s)^2} \frac{\pi^2}{3}
- \frac{3-4\sqrt{s}}{(1+\sqrt{s})^2}
- \frac{1 + 3s}{1 - s} \,\frac{\delta^{1S}(\mu)}{2}
\,, \nn\\
\omega_L^{99}(s) &=
- \sqrt{s}\, \omega^{99}(s) + L(s) + 3 \ln(1-s)
- \frac{8s(1+2s)}{(1-s)^2} \ln(s)
- \frac{5 + s(47-4s)}{2(1-s)}
- \frac{3 + s}{1 - s} \,\frac{\delta^{1S}(\mu)}{2}
\,,\nn\\
\omega^{99}(s) &=
- 4\frac{5+s(12-s)}{(1-s)^2} \Li_2(1-\sqrt{s})
+ \frac{9 - (2-\sqrt{s})^2}{(1-\sqrt{s})^2} \Li_2(1-s)
+ \frac{9 - (2+\sqrt{s})^2}{(1+\sqrt{s})^2} \,\frac{\pi^2}{2}
\,,\\[3ex]
\omega_T^{77}(s) &=
  \frac{\sqrt{s}}{2}\, \omega^{77}(s) + L(s) + 3 \ln(1-s)
- \frac{s(1+5s)}{(1-s)^2} \ln(s)
- \frac{16 + 5s(5-s)}{6(1-s)}
- \frac{3 + s}{1 - s}\, \frac{\delta^{1S}(\mu)}{2}
\,,\nn\\
\omega_L^{77}(s) &=
- \frac{1}{\sqrt{s}}\, \omega^{77}(s) + L(s) + \frac{2+s}{s} \ln(1-s)
+ 2\frac{3 + s(3-s)}{(1-s)^2} \ln(s)
+ \frac{21-s}{2(1-s)}
- \frac{1 + 3s}{1 - s}\, \frac{\delta^{1S}(\mu)}{2}
\,,\nn\\
\omega^{77}(s) &=
  4\frac{12-(3-s)^2}{(1-s)^2} \Li_2(1-\sqrt{s})
- \frac{4 - (1-\sqrt{s})^2}{(1-\sqrt{s})^2} \Li_2(1-s)
- \frac{4 - (1+\sqrt{s})^2}{(1+\sqrt{s})^2} \,\frac{\pi^2}{2}
\,,\\[3ex]
\omega_T^{79}(s) & =
  \frac{1}{2}\, \omega^{79}(s) + L(s) + \frac{1+5s}{2s} \ln(1-s)
- \frac{s(1+3s)}{2(1-s)^2} \ln(s)
- \frac{9-5s}{2(1-s)}
- \frac{1 + s}{1 - s}\, \delta^{1S}(\mu)
\,,\nn\\
\omega_A^{70}(s) & =
  \frac{3 + s(9-2s)}{(1-s)^2} \Li_2(s)
+ \frac{1- s(22-s)}{2s(1-s)} \ln(1-s)
- 2\frac{1 + s(13-4s)}{(1-s)^2} \Li_2(\sqrt{s})
+ \frac{13-3s}{1-s} \ln(1-\sqrt{s})
\nn\\ & \quad
+ \frac{5(1+s)}{2(1-s)^2} \frac{\pi^2}{3}
- \frac{s + (3-2\sqrt{s})^2}{2(1+\sqrt{s})^2}
- \frac{1 + s}{1 - s}\, \delta^{1S}(\mu)
\,,\nn\\
\omega_L^{79}(s) &=
- \omega^{79}(s) + L(s) + \frac{1+2s}{s} \ln(1-s)
+ \frac{s(7-s)}{(1-s)^2} \ln(s)
+ \frac{1+11s}{2(1-s)}
- \frac{1 + s}{1 - s}\, \delta^{1S}(\mu)
\,,\nn\\
\omega^{79}(s) &=
  4\frac{\sqrt{s}(3+s)}{(1-s)^2} \Li_2(1-\sqrt{s})
- \frac{1+\sqrt{s}}{(1-\sqrt{s})^2} \Li_2(1-s)
+ \frac{1-\sqrt{s}}{(1+\sqrt{s})^2}\, \frac{\pi^2}{2}
\,.
\end{align}
\end{widetext}
The functions $\omega^i(s)$ entering $\omega_{T,L}^i(s)$
contain all the dependence on $\sqrt{s}$, which cancels in the $q^2$ spectrum.
All $\ln(\mu/m_b)$ terms that usually appear in the
functions $\omega_i^{77,79}(s)$ have been moved into $\C_7$ (along with the appropriate
constant term contained in $\overline{m}_b/m_b^{1S}$).

The $\chi_i^j(s)$ containing the $\ord(1/m_b^2)$ corrections in Eq.~\eqref{hij_decomp}
can be extracted from Ref.~\cite{Ali:1996bm}:
\begin{align}
\chi_T^{99}(s)
&= - \frac{\lambda_1 + 3\lambda_2}{6}\, \frac{5+3s}{1-s} - 2\lambda_2\, \frac{s(4-3s)}{(1-s)^2}
\,,\nn\\
\chi_A^{90}(s)
&= \frac{\lambda_1 + 3\lambda_2}{6}\, \frac{3+s(2+3s)}{(1-s)^2} - 2 \lambda_2\, \frac{3 + s(4-3s)}{(1-s)^2}
\,,\nn\\
\chi_L^{99}(s)
&= \frac{\lambda_1 + 3\lambda_2}{6}\, \frac{3 + 13s}{1-s} - 2\lambda_2\, \frac{s^2}{(1-s)^2}
\,,\nn\\
\chi_T^{77}(s)
&= \frac{\lambda_1 + 3\lambda_2}{6}\, \frac{3+5s}{1-s} - 2\lambda_2\, \frac{3-2s^2}{(1-s)^2}
\,,\nn\\
\chi_L^{77}(s)
&= - \frac{\lambda_1 + 3\lambda_2}{6}\,\frac{13+3s}{1-s} - 2\lambda_2\, \frac{s(4-3 s)}{(1-s)^2}
\,,\nn\\
\chi_T^{79}(s)
&= \frac{\lambda_1 + 3\lambda_2}{2} - \lambda_2\, \frac{5-3s^2}{(1-s)^2}
\,,\nn\\
\chi_A^{70}(s)
&= \frac{\lambda_1 + 3\lambda_2}{6}\, \frac{3+ s(2+3s)}{(1-s)^2} - \lambda_2\, \frac{5+3s(2-s)}{(1-s)^2}
\,,\nn\\
\chi_L^{79}(s)
&= \frac{\lambda_1 + 3\lambda_2}{2} - 2\lambda_2\, \frac{1}{(1-s)^2}
\,.\end{align}

\section{Numerical Inputs}
\label{app:numeric}

\begin{table}[t]
\tabcolsep 6pt
\begin{tabular}{c|ccc}
\hline\hline
 & $\mu = 2.35\GeV $ & $\mu = 4.7\GeV $ & $\mu = 9.4\GeV $ \\
\hline
$\alpha_s(\mu)$ & 0.2659 & 0.2140 & 0.1793 \\
$C_1(\mu)$ & $-0.4642$ & $-0.2880$ & $-0.1506$ \\
$C_2(\mu)$ &  $1.019$ & $1.007$ & $1.001$ \\
$C_3(\mu)$ & $-0.0096$ & $-0.0043$ & $-0.0017$ \\
$C_4(\mu)$ & $-0.1247$ & $-0.0795$ & $-0.0508$ \\
$C_5(\mu)$ &  $0.00069$ & $0.00029$ & $0.00009$ \\
$C_6(\mu)$ &  $0.00205$ & $0.00081$ & $0.00026$ \\
$C_8(\mu)$ & $-0.2012$ & $-0.1778$ & $-0.1598$ \\
\hline
$\mbbar(\mu)$ & $4.703$ & $4.120$ & $3.707$ \\
$C_7(\mu)$ & $-0.3637$ & $-0.3293$ & $-0.2982$ \\
$\C_7$ & $-0.2435$ & $-0.2611$ & $-0.2687$ \\
\hline
$C_9(\mu)$ & $4.504$ & $4.209$ & $3.790$ \\
$\C_9$ & $4.258$ & $4.207$ & $4.188$ \\
\hline
$\C_{10}$ & $-4.175$ & $-4.175$ & $-4.175$ \\
\hline\hline
\end{tabular}
\caption{\label{Ci_numerical}
Values of the Wilson coefficients to $\ord(\alpha_s)$ at different low scales $\mu$.}
\end{table}

In this Appendix we collect all of our numerical inputs. All values are taken
from Ref.~\cite{pdg} except where stated otherwise. To evaluate the Wilson
coefficients we use
\begin{align}
m_W &= 80.403 \GeV
\,,\nn\\
\sin^2\theta_W &= 0.23122
\,,\nn\\
m_t^\mathrm{pole} &= (171.4 \pm 2.1) \GeV
\,,\nn\\
\alpha_s(m_Z) &= 0.1176
\,,\nn\\
\mu_0^c &= 80 \GeV
\,,\nn\\
\mu_0^t &= 120 \GeV
\,.\end{align}
Here, $\mu_0^{c,t}$ are the matching scales in the charm and top sector,
respectively, and we use the same values as in Ref.~\cite{Bobeth:1999mk}. For
the top-quark mass we use the newest CDF and D0 average~\cite{Brubaker:2006xn}.
The resulting values for the Wilson coefficients at $\ord(\alpha_s)$ run down to
the low scale and the corresponding values for the $\C_i$ according to
Eq.~\eqref{curlyC} are listed in Table~\ref{Ci_numerical}. Note that the
residual scale uncertainties of $\C_7$ and especially $\C_9$ are much smaller
than those of $C_{7,9}(\mu)$.  We use a Mathematica code by Bobeth with the
initial conditions and renormalization group running as given in
Refs.~\cite{Bobeth:1999mk, Bobeth:2003at}. For $C_9(\mu)$ this requires the
three-loop mixings calculated in Refs.~\cite{Gambino:2003zm}.

In the decay rates we use
\begin{align}\label{inputs}
\alpha_\mathrm{em}(m_b) &= 1/133
\,,\nn\\
\lvert V_{tb} V_{ts}^* \rvert &= 41.09 \times 10^{-3}
\,,\nn\\
m_B &= 5.279\GeV
\,,\nn\\
\tau_B &= 1.584\,\mathrm{ps}
\,,\nn\\
m_{K^*} &= 0.892 \GeV
\,,\nn\\
m_b \equiv m_b^{1S} &= (4.70 \pm 0.04)\GeV
\,,\nn\\
m_c \equiv m_c^{1S} &= (1.41 \pm 0.05)\GeV
\,,\nn\\
\lambda_1 &= -0.27 \GeV^2
\,,\nn\\
\lambda_2 &= 0.12 \GeV^2
\,.\end{align}
We use the value of the electromagnetic coupling at the scale $\mu\sim m_b$,
because for the total rate in this case the higher order electroweak corrections
(which we neglect in our analysis) turn out to be numerically small, below the two percent
level~\cite{Bobeth:2003at,Huber:2005ig}. The value of $\lvert V_{tb} V_{ts}^*
\rvert$ is taken from Ref.~\cite{Hocker:2006xb}. For $m_b^{1S}$ we take the
naive average of Refs.~\cite{Bauer:2004ve, Abe:2006xq}, which coincides with the
PDG average~\cite{pdg}, and use the average of the errors quoted in
Refs.~\cite{Bauer:2004ve, Abe:2006xq}. For $m_c^{1S}$ we use the result of
Ref.~\cite{Hoang:2005zw} as quoted in the $1S$ scheme in
Ref.~\cite{Hocker:2006xb}. Finally, for $\lambda_1$ we take the value from
Ref.~\cite{Bauer:2004ve}, and $\lambda_2 = (m_{B^*}^2 - m_B^2)/4 \approx
0.12\GeV^2$.

\end{document}